\documentclass[AMA,Times1COL]{WileyNJDv5} 

\usepackage{comment}
\usepackage{bm}
\usepackage{graphicx}
\usepackage{subcaption}
\articletype{Article Type}%

\received{Date Month Year}
\revised{Date Month Year}
\accepted{Date Month Year}
\journal{Journal}
\volume{00}
\copyyear{2023}
\startpage{1}

\begin{document}

\title{An Empirical Method for Analyzing Count Data}

\author[1]{Jiren Sun}

\author[1]{Linda Amoafo}

\author[1]{Yongming Qu}

\authormark{Sun \textsc{et al.}}
\titlemark{An Empirical Method for Analyzing Count Data}

\address{\orgdiv{Global Statistical Sciences}, \orgname{Eli Lilly and Company}, \orgaddress{\state{Indiana}, \country{USA}}}

\corres{Jiren Sun, \email{jiren.sun@lilly.com}}

\abstract[Abstract]{Count endpoints are common in clinical trials, particularly for recurrent events such as hypoglycemia. When interest centers on comparing overall event rates between treatment groups, negative binomial (NB) regression is widely used because it accommodates overdispersion and requires only event counts and exposure times. However, NB regression can be numerically unstable when events are sparse, and the efficiency gains from baseline covariate adjustment may be sensitive to model misspecification. We propose an empirical method that targets the same marginal estimand as NB regression---the ratio of marginal event rates---while avoiding distributional assumptions on the count outcome. Simulation studies show that the empirical method maintains appropriate Type I error control across diverse scenarios, including extreme overdispersion and zero inflation, achieves power comparable to NB regression, and yields consistent efficiency gains from baseline covariate adjustment. We illustrate the approach using severe hypoglycemia data from the QWINT-5 trial comparing insulin efsitora alfa with insulin degludec in adults with type 1 diabetes. In this sparse-event setting, the empirical method produced stable marginal rate estimates and rate ratios closely aligned with observed rates, while NB regression exhibited greater sensitivity and larger deviations from the observed rates in the sparsest intervals. The proposed empirical method provides a robust and numerically stable alternative to NB regression, particularly when the number of events is low or when numerical stability is a concern.}


\maketitle

\section{Introduction}

Count data are frequently collected in clinical trials, particularly for recurrent events such as hypoglycemia and gastrointestinal episodes in anti-diabetes treatment studies. When the focus is on event processes over time, recurrent-event methods such as the LWYY model\cite{lin2000semiparametric} or frailty models\cite{fu2016hypoglycemic} can be used to compare event rates between treatment groups,\cite{aalen1978nonparametric} and the Nelson-Aalen estimator provides a way to visualize cumulative event rate trends.

In many applications, however, the primary objective is a between-group comparison of overall event rates, typically summarized as a rate ratio. In such settings, individual-level data can be aggregated as counts and analyzed using generalized linear regression approaches, including Poisson or negative binomial (NB) regression. NB regression is often preferred over Poisson regression because it includes an additional parameter to account for overdispersion in the data. Recent research has focused on identifying optimal approaches for NB models and obtaining robust estimates of marginal means.  \cite{luo2013analysis,qu2015estimation,bartlett2018covariate,ye2023robust}

NB regression offers practical advantages over recurrent event methods such as the LWYY model, as it only requires individual-level summary data (event counts and exposure times) without detailed event timing, making it readily applicable to many clinical trial endpoints such as adverse events. However, there are limitations for the NB regression. First, it may fail to converge or may produce estimates that differ substantially from observed event rates when the number of events is small. This can occur when the likelihood surface is relatively flat and numerical optimization has difficulty locating the maximum likelihood estimate. For example, the NB regression failed to produce reasonable results in an analysis of severe hypoglycemia, and aggregated event rates were reported instead. \citep{blevins2016randomized} Second, when adjusting for baseline covariates, the adjusted analysis may be less precise than the unadjusted analysis if the relationship between the baseline covariate(s) and the outcome does not follow a log-linear relationship. For example, the baseline hypoglycemia event rate may be adjusted for the analysis of the number of hypoglycemia events.\citep{luo2013effect} In this case, it may be unrealistic to assume that the baseline hypoglycemia rate is a linear function of logarithm of the postbaseline hypoglycemia event rate because both baseline and postbaseline hypoglycemia are in the same scale or ``unit". 

To overcome the above limitations, we propose an empirical method for analyzing count data with or without adjusting for baseline covariates. The article is arranged as follows. In Section \ref{sec:method}, we present the method. In Section \ref{sec:simulations}, we summarize the simulation settings and simulation results. In Section \ref{sec:example}, we apply the new method to a clinical trial comparing efsitora alfa with insulin degludec in adults with type 1 diabetes. Finally, we provide summary and discussions in Section \ref{sec:discussion}. 

\section{The Empirical Method}\label{sec:method}
\subsection{Estimand}\label{sec:estimand}

Let $Y_{ij}$ denote the count response variable for subject $j=1,2,\ldots,n_i$ in treatment group $i$ ($i=0$ for the control treatment group and $i=1,2,\ldots, I$ for the experimental treatment group). Let $d_{ij}$ denote the exposure or duration of follow-up for subject $j$ in treatment group $i$. Throughout this article, we consider $d_{ij}$'s as fixed quantities, which is consistent with the NB or Poisson regression in which $\log(d_{ij})$ is considered as a fixed offset parameter.

The estimand of interest is the ratio of the marginal event rates between treatment groups. Specifically, we assume that for each treatment group $i$, all subjects share a common constant event rate $r_i$ over time, defined as the expected number of events per unit time in the target population. Under this assumption, $E[Y_{ij}] = d_{ij} r_i$. This assumption defines the parameter of interest (the estimand) but does not impose distributional assumptions on $Y_{ij}$ or require any specific relationship between baseline covariates and outcomes. The rate ratio $\lambda_{ik} = r_i/r_k$ between treatment groups is the primary estimand for treatment comparison. Both the NB regression approach and the empirical method proposed in this article target this same estimand.

\subsection{Inferences}

Let $Y_{i\cdot} = \sum_{j}Y_{ij}$ be the total number of events for treatment group $i$, and $d_{i\cdot} = \sum_{j}d_{ij}$ denote the total exposure or total duration of follow-up for treatment group $i$. The aggregated event rate for treatment group $i$, which is an unbiased estimator for $r_i$, can be expressed as
\begin{equation} \label{eq:rate}
\tilde{r}_{i} = \frac{Y_{i\cdot}}{d_{i\cdot}}, \quad i=0, 1, \ldots, I.
\end{equation}
The aggregated event rate in Equation \eqref{eq:rate} can be re-written as
\begin{equation} \label{eq:rate2}
\tilde{r}_{i} = n_{i}^{-1}\sum\nolimits_{j}W_{ij},
\end{equation}
where $W_{ij} = \left(\bar{d}_{i\cdot}\right)^{-1} Y_{ij}$ and $\bar{d}_{i\cdot} = d_{i\cdot}/n_i$ is the mean exposure or duration of follow-up for treatment group $i$. Equation~\eqref{eq:rate2} effectively transfers the aggregate event rate as the mean of subject-level quantities $W_{ij}$, which are independent across subjects. Therefore, ${r}_{i}$ can be estimated by leveraging the analysis of covariance (ANCOVA) by regressing $W = (W_{11}, W_{12}, \ldots, W_{In_I})'$ on the treatment indicator (as a factor) and baseline covariates $X = (X_{11}, X_{12}, \ldots, X_{In_I})'$, where $X_{ij}$ denotes a vector of baseline covariates for subject $j$ in treatment group $i$. This method works not only for parallel clinical studies but also for cross-over studies. In the analysis of cross-over study, a between-subject random effect is generally included to model the within-subject between-period correlation. \citep{wang2016crossover} Similarly, the more complex Analysis of Heterogeneous Covariance (ANHECOVA) \citep{Ye02102023} can also be used to estimate $r_i$.

Let $\hat{\bm r} = (\hat r_0, \hat r_1, \ldots, \hat r_I)'$ denote the vector of estimators for the event rates obtained from the above linear model, and let $\hat V_{(I+1)\times(I+1)}$ denote the corresponding variance estimator. In most clinical settings, primary interest lies in pairwise comparisons of event rates between treatment arms, which are naturally summarized through the rate ratio
\[
\lambda_{ik} := \frac{r_i}{r_k}, \qquad 0 \le i \neq k \le I.
\]
Accordingly, it is convenient to work on the log scale and define $\theta_i := \log(r_i)$, so that $\log(\lambda_{ik}) = \theta_i - \theta_k$. Let $\hat{\bm \theta} = (\hat \theta_0, \hat \theta_1, \ldots, \hat \theta_I)'$, where $\hat \theta_i = \log(\hat r_i)$. By the $\delta$-method, the variance of $\hat{\bm \theta}$ is given by
\[
\hat V_{\theta} = \left\{\frac{\partial \log(\hat{\bm r})}{\partial \hat{\bm r}} \right\}' \hat V \left\{\frac{\partial \log(\hat{\bm r})}{\partial \hat{\bm r}} \right\}.
\]

Inference for any pairwise comparison can then be conducted directly on the log-rate scale. In particular, testing
\[
H_0: \theta_i = \theta_k
\quad \text{(equivalently, } H_0: r_i = r_k \text{)}
\]
can be based on the $z$-statistic:
\begin{equation*} \label{eq:z_stat}
Z_{ik} = \frac{\hat \theta_i - \hat \theta_k}{\sqrt{\widehat{Var}\left(\hat \theta_i - \hat \theta_k\right)}}.
\end{equation*}
The $100(1-\alpha)\%$ confidence interval (CI) for $\theta_i - \theta_k$ can be constructed using an approximate normal distribution with mean $\hat \theta_i - \hat \theta_k$:
\begin{equation} \label{eq:CI_mu}
\left[ (\hat\theta_i - \hat\theta_k) - z_{1 - \alpha/2}\sqrt{\widehat{Var}\left(\hat \theta_i - \hat \theta_k\right)}, \; (\hat\theta_i - \hat\theta_k) + z_{1 - \alpha/2}\sqrt{\widehat{Var}\left(\hat \theta_i - \hat \theta_k\right)} \right].
\end{equation}
Applying an exponential transformation for the CI in Equation \eqref{eq:CI_mu}, we can obtain the confidence interval for the rate ratio $\lambda_{ik} := r_i/r_k$: 
\begin{equation*} \label{eq:CI_lambda}
\left[ \hat \lambda_{ik} \exp\left\{- z_{1 -\alpha/2}\sqrt{\widehat{Var}\left(\hat \theta_i - \hat \theta_k\right)} \right\}, \; \hat \lambda_{ik} \exp\left\{z_{1 -\alpha/2}\sqrt{\widehat{Var}\left(\hat \theta_i - \hat \theta_k\right)} \right\} \right],
\end{equation*}
where $\hat \lambda_{ik} = \hat r_{i}/\hat r_{k}$.

Note the event rate $r_i$ can also be estimated by the mean of $\ddot{r}_i = Y_{ij}/d_{ij}$. If all subjects have the same duration of follow-up, this estimator is the same as $\tilde r_i$ given in Equation \eqref{eq:rate} or \eqref{eq:rate2}. If the durations of follow-up are different between subjects, which is often the case in clinical trials due to dropouts or administrative censoring, $\ddot{r}_i$ is often subject to larger variability compared to $\tilde{r}_i$ or $\hat r_i$ because there may be some subjects with non-zero number of events in a very short duration of follow-up. Therefore, $\ddot{r}_i$ or any estimator based on a linear model for the response variable $Y_{ij}/d_{ij}$ is not recommended.

A meta-analysis approach or adjusting for strata based on the above proposed empirical estimator is provided in Appendix \ref{sec:Meta}. 

\section{Simulation Studies}\label{sec:simulations}

We conducted two simulation studies to evaluate the operating characteristics of the proposed empirical method in comparison with NB regression. Across both studies, NB regression was implemented using the approach recommended by Luo and Qu (2013),\cite{luo2013analysis,qu2015estimation} which combines sandwich variance estimation with Pearson overdispersion correction, as this approach has been shown to be most robust to model misspecification while maintaining appropriate Type I error control. 

The first simulation study examines Type I error and power under marginal NB count distributions with varying degrees of correlation between baseline and postbaseline counts. The second study investigates performance in the presence of excessive structural zero.

\subsection{Simulation Study 1: Correlated Negative Binomial Counts}
\subsubsection{Data Generation}

In all simulation studies, we consider two treatment groups $i=0$ for the control treatment group and $i=1$ for the experimental treatment group. For each simulated dataset, for treatment group $i$, we generated correlated baseline count data $X_{ij}$ and postbaseline count data $Y_{ij}$, $j=1,2,\ldots, n_i$. The data generation process involved the following steps:
\begin{enumerate}
    \item \textbf{Exposure Time:} For each subject, the postbaseline follow-up duration was independently drawn from one of two uniform distributions, $U(0.6,1.2)$ or $U(0.8,1.4)$, each with probability 0.5. This design yields heterogeneous follow-up times while maintaining an average follow-up duration of approximately 1.
    \item \textbf{Marginal Count Distributions}: Generate count data from NB distribution with $X_{ij} \sim NB(r_x, k_x)$ and $Y_{ij} \sim NB(r_i, k_i)$, where $r_x$ and $k_x$ are the mean and dispersion parameter, respectively, for baseline covariate, and $r_i$ and $k_i$ are the mean and dispersion parameter, respectively, for postbaseline count data for treatment group $i$. 
    \item \textbf{Correlation Structure:} To induce correlation between $X_{ij}$ and $Y_{ij}$, we employed a Gaussian copula approach. Latent bivariate normal variables were generated with a latent correlation, then transformed to uniform margins via the standard normal CDF, and finally converted to NB variates using inverse CDFs with the specified parameters. The latent correlation was calibrated via root-finding to achieve the target observed correlation $\rho$ (the correlation between the realized count variables $X_{ij}$ and $Y_{ij}$).
    \item \textbf{Treatment Effect:} The true event rates in the postbaseline period between two treatment groups were specified as $r_1 = r_0 \exp(\beta_{\text{trt}})$, where$\beta_{\text{trt}}$ is the log rate ratio. Under the null hypothesis, $\beta_{\text{trt}}=0$; under the alternative, $\beta_{\text{trt}}\neq 0$.
\end{enumerate}

\subsubsection{Simulation Scenarios}

Six scenarios (Cases A--F) were considered, reflecting a range of event rates and dispersion levels. The values chosen for these parameters in the simulation studies were informed by realistic hypoglycemia data observed in diabetes studies, particularly for patients with type 1 and type 2 diabetes treated with insulin therapies. Cases A, B, D, and E were designed to assess Type I error under the null hypothesis of no treatment effect, while Cases C and F evaluated power under clinically meaningful alternatives.

\begin{itemize}
    \item \textbf{Case A:} Low baseline rate ($r_x = 0.4$), moderate postbaseline rate ($r_0 = r_1= 0.7$), and high dispersion ($k_x = 3.75$ for baseline, $k_0 = k_1 = 2.43$ for postbaseline). This scenario was designed to mimic nocturnal hypoglycemic event rates for type 1 diabetes patients, which typically occur at lower frequencies than total hypoglycemia but remain clinically important. 
    \item \textbf{Case B:} Low baseline rate ($r_x = 0.4$), moderate postbaseline rate ($r_0 = r_1= 0.5$), and extreme high dispersion ($k_x = 3.75$ for baseline, $k_0 = k_1 = 14.0$ for postbaseline). This represents an extreme dispersion scenario that has been observed in some nocturnal hypoglycemia datasets where a small subset of patients experience many recurrent events while most patients have few or no events. This case tests the robustness of methods under very extreme distributional conditions.
    \item \textbf{Case C:} Low baseline rate with same values for the parameters as Case A except with $r_1=0.5$ and $r_0=0.7$ for the event rates in the treatment and control arms, respectively, yielding a treatment effect (rate ratio) of $0.5/0.7 \approx 0.71$. This represents a clinically meaningful reduction in nocturnal hypoglycemia.
    \item \textbf{Case D:} High baseline rate ($r_x = 3.7$), high postbaseline rate ($r_0=r_1=5.6$), and moderate dispersion ($k_x = 2.02$ for baseline, $k_0=k_1 = 0.62$ for postbaseline). This scenario reflects total hypoglycemic event rates (including both nocturnal and daytime events) in diabetes patients, with means approximating 3.7 and 5.6 events per patient. These higher event rates are typical when considering all hypoglycemic episodes rather than nocturnal events alone.
    \item \textbf{Case E:} High baseline rate ($r_x = 3.7$), high postbaseline rate ($r_0=r_1=5.6$), and high dispersion ($k_x = 2.02$ for baseline, $k_0=k_1 = 3.01$ for postbaseline). Similar to Case D but with increased dispersion for postbaseline, representing scenarios where variability in total hypoglycemia increases during follow-up.   
    \item \textbf{Case F:} High baseline rate with same values for the parameters as Case D except with $r_1=4.7$ and $r_0=5.6$ for the event rates for the treatment and control arms, respectively, yielding a treatment effect (rate ratio) of $4.7/5.6 \approx 0.84$. 
\end{itemize}

For each case, we varied the correlation between baseline and postbaseline event counts, using $\rho \in \{0, 0.25, 0.5, 0.75\}$ to span scenarios from no correlation to extremely strong correlation, and we varied the sample size with $n \in \{400, 1000\}$ to reflect small and moderate diabetes-trial settings.

\subsubsection{Methods Compared}
Four analysis methods were applied to each simulated dataset:

\begin{itemize}
    \item \textbf{Unadjusted NB:} NB regression with the treatment assignment as the only covariate and log of follow-up time as an offset. This method estimates the variance-covariance matrix using the classical sandwich estimator while accounting for overdispersion through the Pearson Chi-square statistic, providing robustness to model misspecification. \citep{luo2013analysis}
    \item \textbf{Adjusted NB:} NB regression with baseline count $X_{ij}$ as covariate, the treatment group as a factor, with logarithm of the follow-up time as an offset, also using sandwich variance estimation with Pearson overdispersion correction. Luo and Qu (2013) demonstrated that this model maintains appropriate Type I error control even when the conditional distribution given baseline covariates departs from the NB form. \cite{luo2013analysis} 
    \item \textbf{Unadjusted Empirical:} The proposed empirical method without covariate adjustment (ANOVA-style analysis on the transformed variable $W_{ij}$).
    \item \textbf{Adjusted Empirical:} The proposed empirical method using ANCOVA with baseline count $X_{ij}$ as a covariate and the treatment group as a factor. 
\end{itemize}

For the empirical methods, inference on $\hat r_i$ was based on robust variance estimators as implemented in the \texttt{RobinCar} package. Importantly, covariate adjustment in both the NB regression and the empirical method does not alter the estimand defined in Section~\ref{sec:estimand}.

\subsubsection{Results}

\begin{figure}[htbp]
  \centering
  \includegraphics[width=1\textwidth]{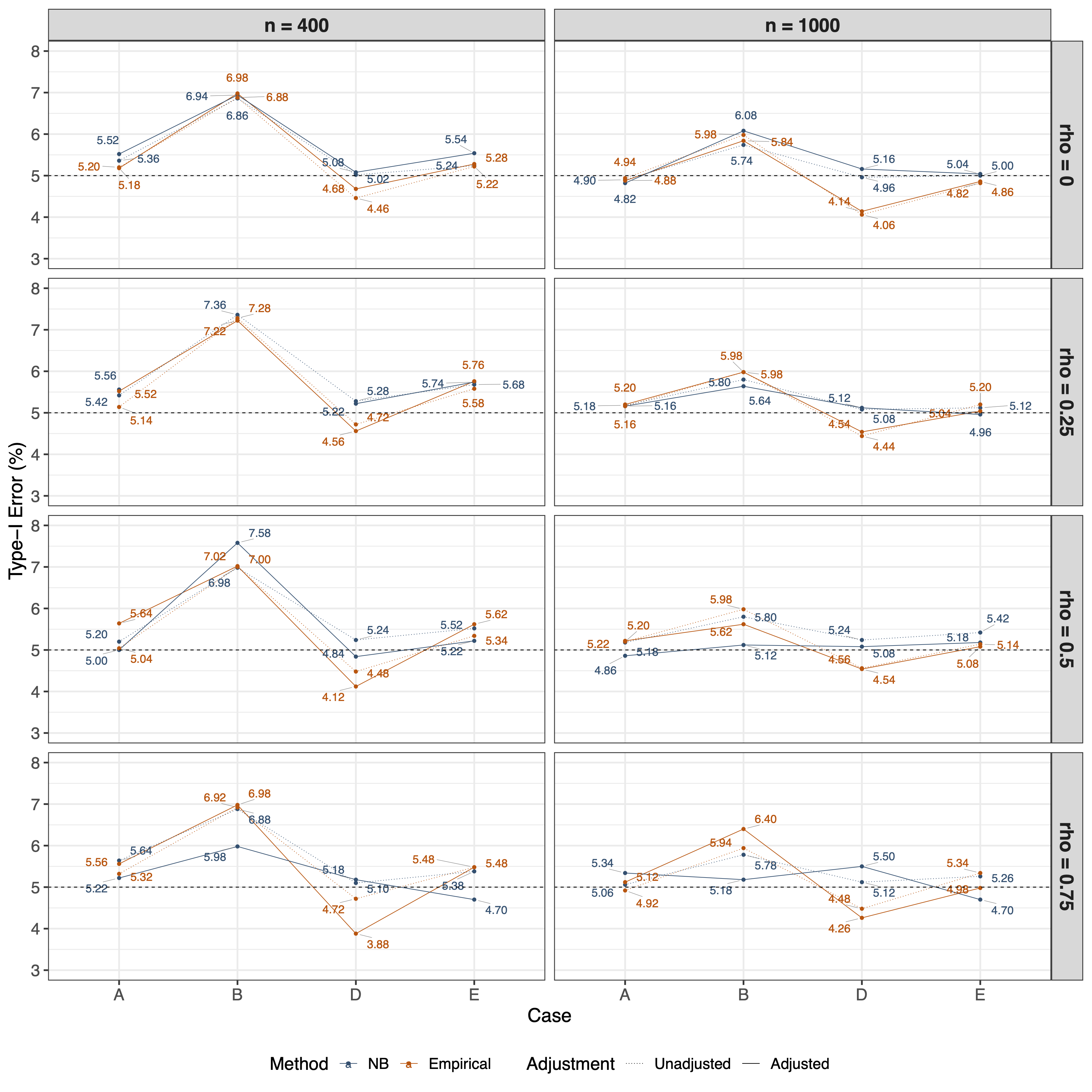}
  \caption{Type I error rates for the null scenarios (Cases A, B, D, E) across different correlations and sample sizes.}
  \label{fig:figure1}
\end{figure}

\begin{figure}[htbp]
  \centering
  \includegraphics[width=1\textwidth]{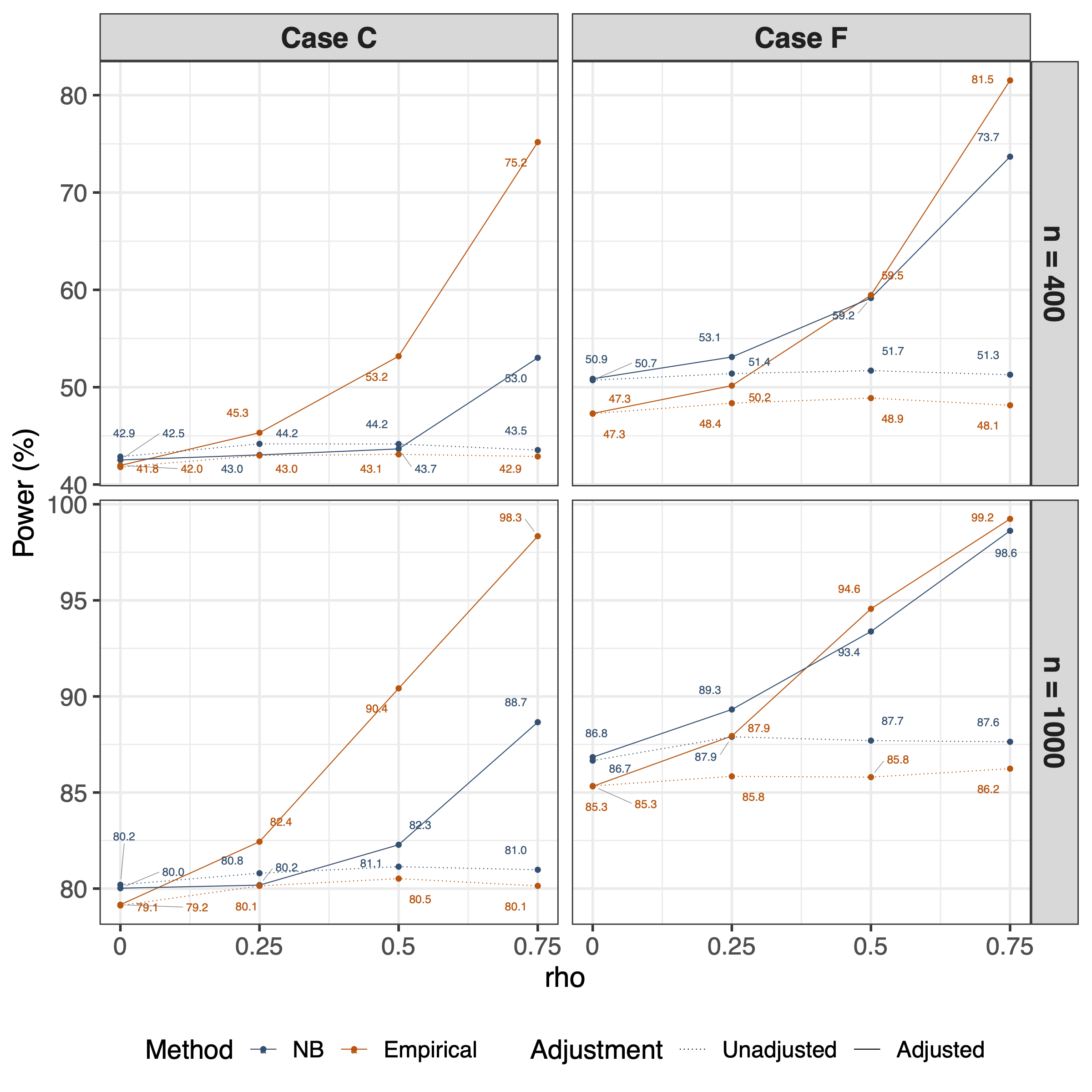}
  \caption{Power for Cases C and F across different correlations and sample sizes.}
  \label{fig:figure2}
\end{figure}

Type I error rates were evaluated at the nominal $\alpha=0.05$ level for null scenarios (Cases A, B, D, E). Power was evaluated for alternative scenarios (Cases C, F) as the probability of rejecting the null hypothesis. Each scenario was replicated 5,000 times. Figure \ref{fig:figure1} displays Type I error rates for the null scenarios (Cases A, B, D, E) across different correlations and sample sizes, while Figure \ref{fig:figure2} shows power for the alternative scenarios (Cases C, F).

All methods maintained appropriate Type I error control near the nominal 5\% level across most scenarios. The empirical method showed no systematic inflation or deflation of Type I error rates when adjusting for baseline covariates. For Cases A, D, and E (moderate-to-high overdispersion scenarios), Type I error rates are close to 5\% for all methods regardless of baseline adjustment or correlation structure. 

Case B (extreme overdispersion with $k_0=k_1 = 14.0$) presented the most extreme scenario. Type I error rates show slight inflation. Notably, both NB and empirical methods showed comparable Type I error rates in this extreme case, with adjusted methods maintaining similar Type I error rates as unadjusted methods. The robustness in Case B demonstrates that both the NB model and the empirical method can handle data with extremely high overdispersion while maintaining acceptable Type I error control.

Figure \ref{fig:figure2} demonstrates the power of NB and empirical methods. Without baseline adjustment, NB showed slightly higher power than the empirical method across scenarios. Substantial power gains were achieved from baseline covariate adjustment when correlation between baseline and follow-up was moderate to high. Notably, the adjusted empirical method achieved higher power than adjusted NB in Case C, contrary to the common expectation that NB regression has superior power. The reason adjusted NB has lower power in Case C may be because conditional on the baseline covariate, the postbaseline count no longer follows a NB distribution, so the NB model is misspecified. In fact, at low to moderate correlations in Case C, adjusted NB even showed slightly lower power than unadjusted NB, demonstrating that there is no guaranteed efficiency gain from covariate adjustment in NB regression: if the model is misspecified, adjusted NB can be less efficient than unadjusted NB. In contrast, for the empirical method, adjusted estimator consistently showed higher power than unadjusted estimator across all correlation levels, demonstrating consistent efficiency gains with baseline covariate adjustment. For Case F, the two adjusted estimators exhibited comparable power.

\subsection{Simulation Study 2: Zero-Inflated Negative Binomial Counts}
\subsubsection{Data Generation}

In Simulation Study~1, excess zero counts arise naturally in each treatment arm as a consequence of low event rates and overdispersion. The second simulation study considers a different data-generating mechanism by explicitly introducing a structural zero component through a zero-inflated negative binomial (ZINB) model. In this setting, a subset of participants is assumed to be unable to experience events during follow-up, resulting in structural zeros. The data generation process was as follows:
\begin{enumerate}
    \item \textbf{Baseline Covariates:} Let $X_{ij}$ denote the baseline event count, assumed to follow a Poisson distribution with mean 1.5. Let $Z_{ij}$ denote a continuous covariate from $N(0, 1)$, representing a standardized biomarker or prognostic factor.
    \item \textbf{Exposure Time:} Generated identically to Simulation Study 1.
    \item \textbf{Zero-Inflated Negative Binomial Model:} The follow-up count $Y_{ij}$ was generated from a ZINB distribution. With probability $\pi$ (the zero-inflation probability), $Y_{ij} = 0$ (a structural zero representing subjects who cannot experience events). With probability $(1 - \pi)$, $Y_{ij}$ follows a NB distribution with mean $\mu_i$ and variance $\mu_i + k_i\mu_i^2$, where $i = 0$ for control and $i = 1$ for treatment, $\mu_i = \beta_0 \exp(\beta_{\text{trt}} i + \beta_1 X_{ij} + \beta_2 Z_{ij})\cdot d_{ij}$, and $k_0 = k_1 = 1$ is the dispersion parameter.
    \item \textbf{Treatment Effect:} $\beta_{\text{trt}}=0$ for null scenarios, and $\beta_{\text{trt}}=\log(0.7)$ for alternative scenarios.
\end{enumerate}

\subsubsection{Simulation Scenarios}

We considered four scenarios (Cases G--J). Under the null hypothesis, Case G assumes high zero-inflation ($\pi = 0.6$, meaning 60\% structural zeros) with a low event rate among non-zero subjects ($\beta_0 = \log(0.3)$), while Case H assumes moderate zero-inflation ($\pi = 0.3$) with the same low event rate. Under the alternative hypothesis, Case I mirrors Case G but incorporates a treatment effect with $\beta_{\text{trt}}=\log(0.7)$, and Case J mirrors Case H with the same treatment effect. For each case, the sample size was $n \in \{400, 1000\}$, and the covariate effects were set to relatively large values $\beta_1 = \log(1.5)$ and $\beta_2 = \log(2)$ to make $X_{ij}$ and $Z_{ij}$ strong predictors of the follow-up count.

\subsubsection{Results}

Each scenario was replicated 5,000 times. For Cases G and I, the control arm had a mean follow-up count of approximately 0.32 events per patient ($SD=1.17$). For Cases H and J, the control arm had a mean of approximately 0.56 events ($SD=1.52$). The treatment arms in Cases I and J showed reductions to 0.23 events ($SD=0.86$) and 0.40 events ($SD=1.12$), respectively. The observed correlations between baseline and postbaseline counts were moderate ($\rho \approx 0.16$-$0.22$), and the continuous covariate $Z_{ij}$ showed similar correlations with $Y_{ij}$ ($\rho \approx 0.19$-$0.26$), confirming that both covariates were meaningful predictors of follow-up event rates.

\begin{figure}[htbp]
  \centering
  \includegraphics[width=0.7\textwidth]{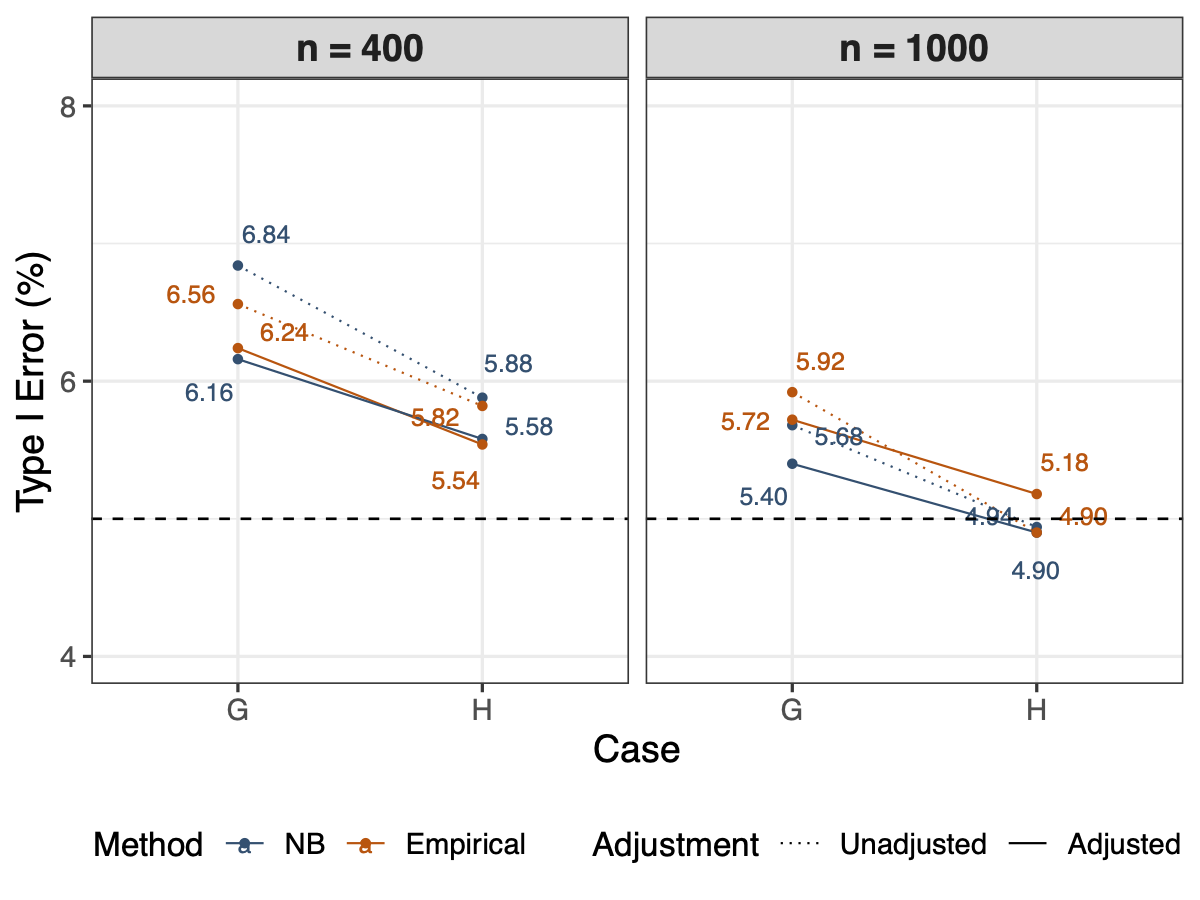}
  \caption{Type I error rates for the null scenarios (Cases G and H) across different correlations and sample sizes.}
  \label{fig:figure3}
\end{figure}

\begin{figure}[htbp]
  \centering
  \includegraphics[width=0.7\textwidth]{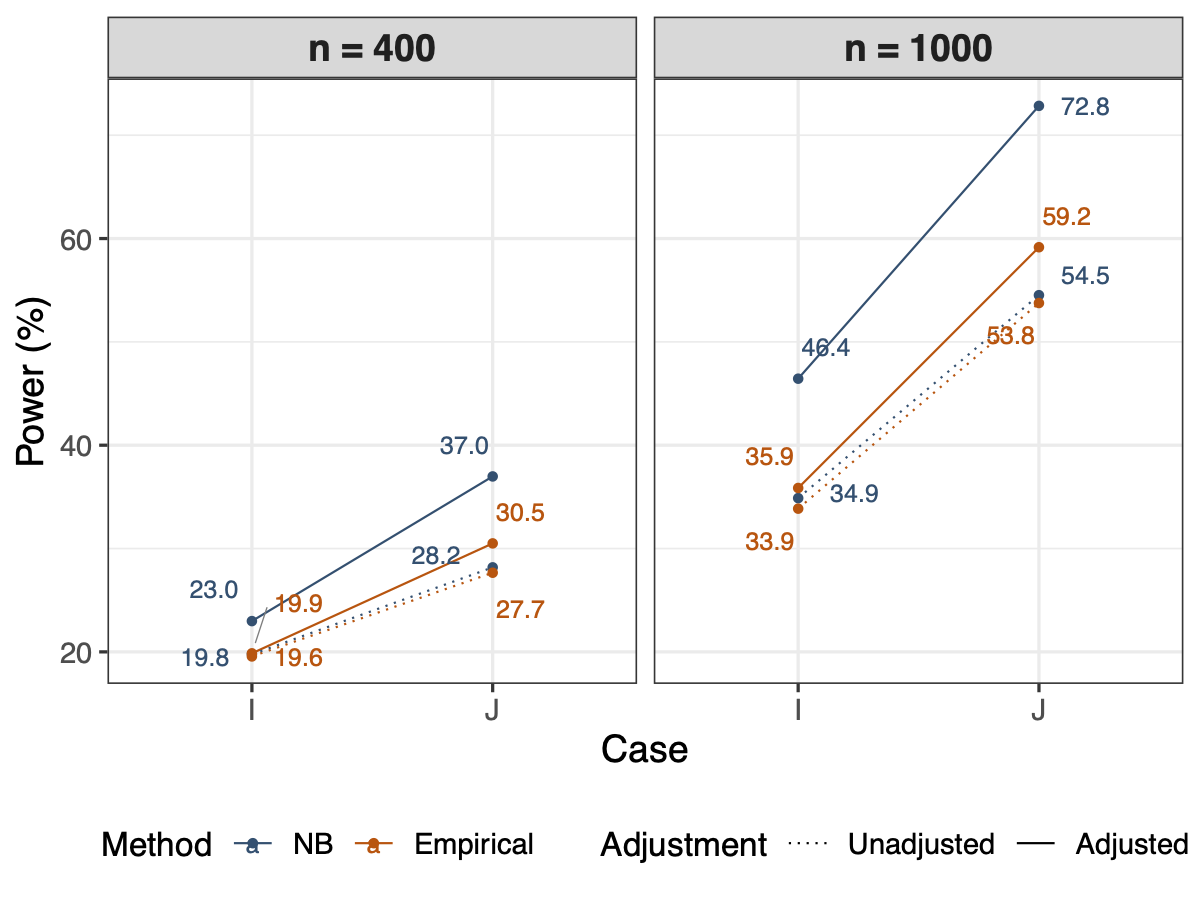}
  \caption{Power for Cases I and J across different correlations and sample sizes.}
  \label{fig:figure4}
\end{figure}

The same four methods as Simulation Study 1 were applied. Figure \ref{fig:figure3} displays Type I error rates for the null scenarios (Cases G and H), and Figure \ref{fig:figure4} shows power for the alternative scenarios (Cases I and J).

Both the NB and empirical methods maintained appropriate Type I error control in the presence of zero-inflation. For Case G (high zero-inflation, $\pi = 0.6$), Type I error was slightly inflated. Although neither method explicitly models the zero-inflation mechanism, both generally preserved Type I error.

In Figure 4, the NB and empirical methods exhibited nearly identical power across all scenarios when baseline adjustment was not applied. Substantial gains in power emerged once baseline covariates were included. The correlations between baseline counts and postbaseline counts were moderate ($\rho \approx 0.16$-$0.22$), and the continuous covariate $Z_{ij}$ showed similar correlations with postbaseline counts ($\rho \approx 0.19$-$0.26$), which explains the sizable improvements in power with covariate adjustment. With adjustment, the NB method consistently outperformed the empirical method. This is expected because the data were generated from an NB model (aside from structural zeros), the covariates were included in the correct parametric form, and they were both strong predictors of the outcome.

\subsection{Summary of Simulation Findings}
Both simulation studies demonstrate that the proposed empirical method offers a robust and efficient alternative to NB regression for analyzing count data in clinical trials. The empirical method consistently maintained appropriate Type I error rates across all scenarios, including extreme overdispersion (Case B with $k_0=k_1 = 14$) and high zero-inflation (Case G with $\pi = 0.6$). Baseline covariate adjustment did not inflate Type I error for the empirical method. The NB regression also maintained good Type I error control, validating its use as the comparison standard. 

Regarding power, baseline covariate adjustment substantially improved power when baseline-follow-up correlation was moderate to high. In Simulation Study 1, the empirical method showed comparable or even higher power than NB regression in Case C. Notably, at low to moderate correlations in Case C, adjusted NB showed slightly lower power than unadjusted NB, demonstrating that covariate adjustment in NB regression does not guarantee efficiency gains under model misspecification. In contrast, adjusted empirical method consistently outperformed unadjusted empirical across all correlation levels. In Simulation Study 2, NB regression showed a power advantage when adjusted for baseline, which is expected since the generating model was NB (aside from structural zeros) with covariates entered in the correct parametric form as strong predictors. Without baseline adjustment, both methods showed similar power across scenarios.

\section{Real Data Example}\label{sec:example}
\subsection{Study Design and Data}

We applied the empirical method to data from QWINT-5, a phase 3, multicenter, randomized, parallel-design, open-label study comparing insulin efsitora alfa, a novel once-weekly basal insulin, with insulin degludec in adults with type 1 diabetes treated with multiple daily injection therapy.\cite{bergenstal2024once} The study randomized 692 participants in a 1:1 ratio to receive either insulin efsitora alfa or insulin degludec for 52 weeks of treatment. The study was conducted in accordance with the Declaration of Helsinki guidelines on good clinical practices (World Medical
Association 2010).\cite{world2010world}

For this analysis, we focused on severe hypoglycemia events (Level 3 hypoglycemia), defined as episodes requiring assistance for treatment. We selected severe hypoglycemia because the small number of events provides an ideal setting for comparing the numerical stability of the empirical and NB methods when data are sparse.

\subsection{Statistical Methods}

We compared results from NB regression with the empirical method for analyzing severe hypoglycemia event rates across four time intervals: Week 0--12, Week 12--26, Week 26--52, and the overall treatment period (Week 0--52). For both approaches, the response variable was the number of severe hypoglycemia events per participant during each interval.

For the adjusted analyses, both methods included treatment assignment, baseline HbA1c, and baseline severe hypoglycemia rate (events per year during the lead-in period) as covariates. The NB regression used $\log(\text{exposure}/365.25)$ as an offset while the empirical method transformed the count data to $W_{ij} = Y_{ij}/\bar{d}_i$ (where $Y_{ij}$ is the event count for participant $j$ in treatment group $i$ and $\bar{d}_i$ is the mean exposure time in treatment group $i$) and performed ANCOVA on $W_{ij}$ with the same covariate adjustments. For the unadjusted analyses, only treatment assignment was included, with the offset retained for NB regression. For both analyses, rate ratios and 95\% CIs were calculated on the log scale using the delta method and back-transformed. 

\subsection{Results}

Table \ref{tab:severe_hypo_rates_both} presents the analysis results for severe hypoglycemia across different time intervals. The insulin efsitora alfa arm experienced 44 events over 325 patient-years during the overall treatment period (observed rate: 0.135 events/patient/year), while the degludec arm experienced 13 events over 330 patient-years (observed rate: 0.039 events/patient/year), yielding a raw rate ratio (RR) of 3.44. Event counts varied across intervals, with the Week 0--12 period showing 28 vs 7 events (raw RR = 4.05), Week 12--26 showing 7 vs 4 events (raw RR = 1.78), and Week 26--52 showing 9 vs 2 events (raw RR = 4.55).

\begin{table}[htbp]
\centering
\caption{Severe hypoglycemia event rates and rate ratios from negative binomial and empirical methods: unadjusted and adjusted analyses.}
\label{tab:severe_hypo_rates_both}
\footnotesize
\setlength{\tabcolsep}{3.5pt}

\begin{subtable}{\linewidth}
\centering
\caption{Unadjusted Analysis}
\label{tab:severe_hypo_rates_unadjusted}
\renewcommand{\arraystretch}{1.25}
\begin{tabular}{lccccccccccccc}
\toprule
& \multicolumn{5}{c}{Efsitora} 
& \multicolumn{5}{c}{Degludec} 
& \multicolumn{3}{c}{Efsitora vs Degludec} \\
\cmidrule(lr){2-6} \cmidrule(lr){7-11} \cmidrule(lr){12-14}
Period
& Ev & Exp & Obs & NB & Emp
& Ev & Exp & Obs & NB & Emp
& Raw RR
& NB RR (95\% CI)
& Emp RR (95\% CI) \\
\midrule
Week 0--12
& 28 & 78.5 & 0.357 & 0.356 & 0.357
& 7  & 79.6 & 0.088 & 0.088 & 0.088
& 4.05
& 4.05 (1.68, 9.74)
& 4.05 (1.60, 10.26) \\
Week 12--26
& 7  & 88.1 & 0.080 & 0.080 & 0.080
& 4  & 89.8 & 0.044 & 0.044 & 0.044
& 1.78
& 1.80 (0.42, 7.73)
& 1.78 (0.44, 7.24) \\
Week 26--52
& 9  & 159  & 0.057 & 0.057 & 0.057
& 2  & 161  & 0.012 & 0.012 & 0.012
& 4.55
& 4.55 (0.98, 21.06)
& 4.55 (0.99, 20.95) \\
Week 0--52
& 44 & 325  & 0.135 & 0.137 & 0.135
& 13 & 330  & 0.039 & 0.039 & 0.039
& 3.44
& 3.49 (1.76, 6.94)
& 3.44 (1.64, 7.19) \\
\bottomrule
\end{tabular}
\end{subtable}

\vspace{2mm}

\begin{subtable}{\linewidth}
\centering
\caption{Adjusted Analysis}
\label{tab:severe_hypo_rates_adjusted}
\renewcommand{\arraystretch}{1.25}
\begin{tabular}{lccccccccccccc}
\toprule
& \multicolumn{5}{c}{Efsitora} 
& \multicolumn{5}{c}{Degludec} 
& \multicolumn{3}{c}{Efsitora vs Degludec} \\
\cmidrule(lr){2-6} \cmidrule(lr){7-11} \cmidrule(lr){12-14}
Period
& Ev & Exp & Obs & NB & Emp
& Ev & Exp & Obs & NB & Emp
& Raw RR
& NB RR (95\% CI)
& Emp RR (95\% CI) \\
\midrule
Week 0--12
& 28 & 78.5 & 0.357 & 0.359 & 0.358
& 7  & 79.6 & 0.088 & 0.087 & 0.086
& 4.05
& 4.14 (1.71, 10.00)
& 4.16 (1.62, 10.67) \\
Week 12--26
& 7  & 88.1 & 0.080 & 0.094 & 0.086
& 4  & 89.8 & 0.045 & 0.037 & 0.038
& 1.78
& 2.51 (0.56, 11.16)
& 2.29 (0.55, 9.45) \\
Week 26--52
& 9  & 159  & 0.057 & 0.083 & 0.060
& 2  & 161  & 0.012 & 0.009 & 0.009
& 4.55
& 8.76 (1.15, 66.62)
& 6.40 (1.09, 37.64) \\
Week 0--52
& 44 & 325  & 0.135 & 0.152 & 0.139
& 13 & 330  & 0.039 & 0.036 & 0.036
& 3.44
& 4.23 (2.06, 8.67)
& 3.92 (1.82, 8.45) \\
\bottomrule
\end{tabular}
\end{subtable}

\vspace{1.5mm}
\begin{minipage}{0.98\linewidth}
\footnotesize
\textit{Ev} = number of severe hypoglycemia events; 
\textit{Exp} = total exposure time in patient-years; 
\textit{Obs} = observed event rate (events per patient-year); 
\textit{NB} = estimated marginal rate from negative binomial regression; 
\textit{Emp} = estimated marginal rate from the empirical method; 
\textit{Raw RR} = ratio of observed event rates (efsitora/degludec); 
\textit{NB RR} = rate ratio estimated from the negative binomial regression; 
\textit{Emp RR} = rate ratio estimated from the empirical method.
\end{minipage}
\end{table}

Without baseline covariate adjustment, the empirical method's estimated marginal event rates exactly matched the observed rates across all time intervals because the formula for calculation of RR is the same for both estimates. The marginal event rates estimated from NB regression were very close to the observed rates across most intervals, although small discrepancies were present. Despite these differences in marginal rate estimates, the RRs obtained from the NB and empirical methods were similar across all time intervals.

With baseline covariate adjustment, the empirical method’s estimated marginal event rates remained close to the observed values across all intervals. The adjusted NB model also produced reasonable marginal rate estimates overall; however, the deviations from the observed rates were generally larger than those from the empirical method. This difference was most evident in the Week~26--52 interval, where event counts were particularly sparse. In this interval, the marginal RR estimated from NB regression deviated markedly from the raw RR, while the empirical method produced a more stable estimate that lay between the raw RR and the NB estimate. 

This real data example highlights the practical advantages of the empirical method for analyzing rare events. Although NB regression performed reasonably well in this analysis, especially in intervals with more events, the empirical method consistently yielded marginal rate estimates and RRs that more closely tracked the observed data across time intervals and adjustment strategies. This discrepancy arises because NB regression's maximum likelihood estimation can be unstable with sparse counts. The empirical method’s ability to provide stable and interpretable marginal rate estimates that closely reflect observed data is particularly valuable when accurate characterization of event rates in each treatment arm is essential, such as in regulatory submissions for safety endpoints.

Finally, in this example, the unadjusted empirical analysis produced marginal event rate estimates and RRs that were closer to the observed (raw) marginal means and raw RRs than the adjusted empirical analysis. This behavior is not unexpected in settings with extremely sparse events, such as severe hypoglycemia, where baseline event rates are themselves rare and highly variable. In such cases, baseline covariates may offer limited stable prognostic information, and covariate adjustment can introduce additional variability without yielding meaningful gains in precision. As a result, for datasets characterized by very low event counts and noisy baseline rates, unadjusted analyses may be preferable for estimating marginal event rates and RRs in practice.

\section{Discussion}\label{sec:discussion}

We have proposed an empirical method for analyzing count data in clinical trials that targets marginal event rates and rate ratios without relying on a fully specified parametric count model. By transforming event counts to subject-level rate quantities and applying ANCOVA or ANHECOVA with robust variance estimation, the method provides valid inference under minimal assumptions and offers a robust and stable alternative to the NB regression, particularly in settings with sparse events or uncertain model specification.

Through extensive simulation studies, we demonstrated that the empirical method maintains appropriate Type I error control across a wide range of scenarios, including settings with extreme overdispersion and explicit zero inflation. Baseline covariate adjustment did not compromise Type I error control for the empirical method, even when baseline and postbaseline counts were strongly correlated. These findings highlight a key advantage of the empirical approach: its validity does not depend on correct specification of a conditional mean model, nor does it require assumptions about the functional form linking baseline covariates to postbaseline event rates.

Our results also clarify the nuanced role of covariate adjustment in count data analysis. In Simulation Study~1, where the data-generating mechanism did not follow a conditional NB model, covariate adjustment in NB regression did not provide efficiency gains and, in some settings, reduced power relative to unadjusted analyses. In contrast, the empirical method consistently benefited from baseline adjustment across correlation levels, reflecting its model-robust nature. In Simulation Study~2, where data were generated from a zero-inflated NB model with correctly specified covariate effects, adjusted NB regression achieved higher power, as expected under correct model specification (aside from structural zeros). Together, these results emphasize that the efficiency advantages of parametric models depend critically on the correctness of their assumptions, whereas the empirical method offers more stable performance when such assumptions are uncertain.

The real data example from the QWINT-5 trial further illustrates the practical implications of these findings. For severe hypoglycemia, where event counts were extremely sparse, NB regression produced marginal rate estimates that showed larger deviations than the empirical method, particularly after baseline covariate adjustment. In contrast, the empirical method yielded marginal rate estimates and rate ratios that tracked the observed data across time intervals with greater numerical stability. Moreover, in this particular sparse-event setting, unadjusted empirical analyses produced estimates closer to the observed marginal means and raw rate ratios than adjusted analyses. This behavior is not unexpected in settings with very rare events, where baseline event rates are themselves sparse and noisy, limiting their prognostic value and potentially inflating variance when included as covariates. These results suggest that, in some settings, unadjusted analyses may be preferable in practice for estimating marginal event rates when events are extremely rare and baseline covariates provide limited stable information.

In the real data example, the marginal event rate estimates from the NB regression are obtained through the G-computation approach.\citep{qu2015estimation} Recently, augmented inverse propensity weighted (AIPW) estimators have been advocated for obtaining marginal estimates from generalized linear models, including NB regression, which can mitigate potential bias when the working model is misspecified.\citep{bannick2025general} In Appendix~\ref{sec:aipw}, we compare the G-computation and AIPW approaches for the QWINT-5 data and show that both estimators yield similar marginal rate estimates. The primary objective of our proposed empirical method is, however, not to address potential bias in NB regression's marginal estimates, but rather to provide a fundamentally more numerically stable alternative to NB regression itself. In sparse-event settings where maximum likelihood estimation can fail to converge or produce unstable estimates, the empirical method circumvents these issues by avoiding the likelihood-based estimation framework altogether. While AIPW-based corrections could improve marginal estimates from NB regression when it converges, our method addresses the more fundamental concern of estimation stability.

From a regulatory and clinical perspective, the ability to produce stable and interpretable marginal event rate estimates is particularly important for safety endpoints, such as severe hypoglycemia, where accurate characterization of absolute and relative risks informs benefit-risk assessment. The empirical method directly targets these marginal quantities and avoids numerical instability associated with likelihood-based estimation in sparse-data settings, making it well suited for regulatory reporting and clinical interpretation.

In real applications where some baseline covariates are considered to be correlated with the postbaseline count outcome, one needs to take the following considerations in determining whether the NB or the empirical method should be used. As discussed earlier, if events are sparse, the empirical method may be more robust to use. Second, one may consider the relationship between the baseline covariate(s) and the postbaseline count outcome. If one thinks it is likely that the relationship is log-linear, NB regression may be considered; if it is unlikely that the baseline covariate(s) and the count are log-linear, the empirical method may offer an advantage. In the real data example we illustrated in Section \ref{sec:example}, it is unlikely that the baseline hypoglycemia event rate and baseline HbA1c have a linear relationship with the logarithm of the postbaseline event rate. Therefore, the empirical method may be more appropriate for analyzing hypoglycemia events.

Several extensions and limitations merit discussion. First, while the empirical method readily accommodates baseline covariate adjustment, it does not exploit potential efficiency gains from correctly specified parametric models. When the NB model (or a zero-inflated variant) is correctly specified and covariates enter the model in the appropriate functional form, NB regression can achieve modest gains in efficiency, as observed in Simulation Study~2. Second, although our simulations focused on parallel-group trials, the method naturally extends to crossover designs and other settings through standard mixed-model or generalized estimating equation frameworks. Third, in stratified analyses or meta-analyses across multiple studies or disease severity strata, additional considerations arise. When there are multiple strata or multiple studies, event rates may differ across strata (such as ordinal disease severities or individual studies), while the rate ratios are assumed to be the same (or at least similar). Unlike negative binomial regression, which assumes additivity on the logarithmic scale of the event rate across strata, ANCOVA or ANHECOVA analyses may not yield the most efficient estimator for stratified data. In this case, a standard meta-analysis performed directly on the logarithm of the rate ratio across strata, with weights proportional to the total exposure within each stratum, is recommended; a detailed discussion of this setting is provided in the Appendix \ref{sec:Meta}. Fourth, the empirical method presented in this paper utilized ANCOVA for analysis. In practice, clinical trials often employ stratified randomization to ensure balance across important baseline characteristics. The precision of estimates can be enhanced by accounting for the randomization scheme. The empirical method readily accommodates this through the use of ANHECOVA, as implemented in the \texttt{RobinCar} package. Appendix \ref{sec:anhecova} illustrates the application of ANHECOVA to the QWINT-5 data.

In summary, we have developed an empirical approach for analyzing count outcomes in clinical trials that targets marginal event rates and rate ratios under minimal modeling assumptions. The method maintains appropriate Type I error control across a wide range of settings, achieves efficiency comparable to NB regression, and asymptotically benefits from covariate adjustment without relying on correct specification of a conditional mean model. By avoiding likelihood-based estimation, the empirical method offers enhanced numerical stability in sparse-event settings and produces marginal rate estimates that are directly interpretable and closely aligned with observed data. NB regression with robust variance estimation remains a useful and efficient tool when its assumptions are approximately satisfied and event counts are sufficiently large. However, in many practical applications---particularly rare-event settings, the empirical method provides a more robust and stable alternative. Together, these approaches offer complementary strengths, and the choice between them should be guided by the underlying data structure, the plausibility of model assumptions, and the primary scientific objectives of the analysis.

\bmsection*{Acknowledgments}
We thank Michael Case for his insight on the originally planned statistical analysis for the hypoglycemia events in QWINT-5 Study. 

\bmsection*{Ethics}
The study was conducted in accordance with the Declaration of Helsinki guidelines on good clinical practices (World Medical
Association 2010).\cite{world2010world}

\bmsection*{Financial disclosure}
All authors conducted this work as part of their employment, and no external funding was received for this research.

\bmsection*{Conflict of interest}
All authors are employees and minor shareholders of Eli Lilly and Company.

\bibliography{wileyNJD-AMA}

\bmsection*{Supporting information}

Additional supporting information may be found in the
online version of the article at the publisher’s website.

\appendix

\bmsection{Meta-analyis for the rate ratio across strata or studies}\label{sec:Meta}
\vspace*{12pt}

We use additional subscript ``$s$'' to index the stratum (or study). The estimator of the event rate ratio $\lambda_{ik}^{s}$ for treatment group $i$ over treatment group $k$ across strata is
\begin{equation*}
\hat{\lambda}_{ik}^s = \frac{\sum_{s}w_{s}\hat{\lambda}_{ik,s}}{\sum_{s}w_{s}}, 
\end{equation*}
where $\hat{\lambda}_{ik, s}$ is the estimated event rate ratio for treatment group $i$ over treatment group $k$ in stratum $s$, and $w_{s}$ is the total duration of follow-up for all subjects in stratum $s$. 
The variance estimator for $\hat{\lambda}_{ik}^s$ is
\begin{equation*}
\widehat{\text{Var}}(\hat{\lambda}_{ik}^s) = \frac{\sum_{s}w_{s}^{2}\widehat{\text{Var}}(\hat{\lambda}_{ik,s})}{\left( \sum_{k}w_{s} \right)^{2}}.
\end{equation*}

\bmsection{Application of ANHECOVA to the QWINT-5 data}\label{sec:anhecova}
\vspace*{12pt}

In the QWINT-5 trial, participants were randomized in a 1:1 ratio using a stratified randomization scheme. The stratification factors were: (1) country, (2) baseline HbA1c stratum (<8\% or $\geq$8\% at screening), (3) continuous glucose monitoring use prior to study entry (yes/no), and (4) carbohydrate counting for prandial insulin dosing (yes/no). To account for the stratified randomization scheme, we applied ANHECOVA as implemented in the \texttt{RobinCar} package. The model form is the same as before, but the variance calculation now accounts for the stratified randomization scheme.

Table~\ref{tab:anhecova_comparison} presents the rate ratio estimates and 95\% confidence intervals comparing ANCOVA (which does not account for the stratified randomization scheme) with ANHECOVA (which accounts for the stratified randomization scheme). The marginal rate estimates remain identical between methods (not shown), but confidence intervals narrow when the stratified randomization scheme is accounted for, reflecting improved precision. This demonstrates that the empirical method can leverage the stratified randomization scheme to achieve meaningful gains in statistical efficiency.

\begin{table}[htbp]
\centering
\caption{Comparison of confidence intervals for severe hypoglycemia rate ratios: ANCOVA versus ANHECOVA.}
\label{tab:anhecova_comparison}
\renewcommand{\arraystretch}{1.25}
\begin{tabular}{lcccccccc}
\toprule
& & \multicolumn{3}{c}{Unadjusted} & \multicolumn{3}{c}{Adjusted} \\
\cmidrule(lr){3-5} \cmidrule(lr){6-8}
Period & Raw RR & RR & ANCOVA 95\% CI & ANHECOVA 95\% CI & RR & ANCOVA 95\% CI & ANHECOVA 95\% CI \\
\midrule
Week 0--12  & 4.05 & 4.05 & (1.60, 10.26) & (1.66, 9.88)  & 4.16 & (1.62, 10.67) & (1.69, 10.27) \\
Week 12--26 & 1.78 & 1.78 & (0.44, 7.24)  & (0.47, 6.82)  & 2.29 & (0.55, 9.45)  & (0.58, 8.95) \\
Week 26--52 & 4.55 & 4.55 & (0.99, 20.95) & (1.02, 20.25) & 6.40 & (1.09, 37.64) & (1.12, 36.43) \\
Week 0--52  & 3.44 & 3.44 & (1.64, 7.19)  & (1.69, 7.00)  & 3.92 & (1.82, 8.45)  & (1.86, 8.26) \\
\bottomrule
\end{tabular}

\vspace{2mm}
\begin{minipage}{0.95\linewidth}
\footnotesize
\textit{Raw RR} = ratio of observed event rates (efsitora/degludec); \textit{RR} = rate ratio estimated from the empirical method; \textit{ANCOVA} = standard analysis of covariance without accounting for stratified randomization scheme; \textit{ANHECOVA} = analysis of heterogeneous covariance accounting for stratified randomization scheme. Unadjusted analysis includes only treatment assignment and stratification factors; adjusted analysis additionally includes baseline HbA1c and baseline severe hypoglycemia rate as covariates.
\end{minipage}
\end{table}

\bmsection{Comparison of G-Computation and AIPW Estimators for NB Models}\label{sec:aipw}

The marginal rates presented in Table~\ref{tab:severe_hypo_rates_both} under the ``NB'' columns correspond to the G-computation estimator. An alternative approach is the augmented inverse propensity weighted (AIPW) estimator. Table~\ref{tab:severe_hypo_gcomp_aipw} compares the G-computation and AIPW estimators for both unadjusted and adjusted NB regression analyses. The results from these two estimators are similar and close to each other across all time periods.

\begin{table}[htbp]
\centering
\caption{Comparison of G-computation and AIPW estimators for severe hypoglycemia event rates from NB regression.}
\label{tab:severe_hypo_gcomp_aipw}
\footnotesize
\setlength{\tabcolsep}{7pt}
\begin{subtable}{\linewidth}
\centering
\caption{Unadjusted Analysis}
\label{tab:gcomp_aipw_unadjusted}
\renewcommand{\arraystretch}{1.25}
\begin{tabular}{lcccccc}
\toprule
& \multicolumn{3}{c}{Efsitora} 
& \multicolumn{3}{c}{Degludec} \\
\cmidrule(lr){2-4} \cmidrule(lr){5-7}
Period
& Obs & G-comp & AIPW
& Obs & G-comp & AIPW \\
\midrule
Week 0--12
& 0.357 & 0.356 & 0.357
& 0.088 & 0.088 & 0.088 \\
Week 12--26
& 0.080 & 0.080 & 0.079
& 0.044 & 0.044 & 0.045 \\
Week 26--52
& 0.057 & 0.057 & 0.057
& 0.012 & 0.012 & 0.012 \\
Week 0--52
& 0.135 & 0.137 & 0.135
& 0.039 & 0.039 & 0.039 \\
\bottomrule
\end{tabular}
\end{subtable}
\vspace{2mm}
\begin{subtable}{\linewidth}
\centering
\caption{Adjusted Analysis}
\label{tab:gcomp_aipw_adjusted}
\renewcommand{\arraystretch}{1.25}
\begin{tabular}{lcccccc}
\toprule
& \multicolumn{3}{c}{Efsitora} 
& \multicolumn{3}{c}{Degludec} \\
\cmidrule(lr){2-4} \cmidrule(lr){5-7}
Period
& Obs & G-comp & AIPW
& Obs & G-comp & AIPW \\
\midrule
Week 0--12
& 0.357 & 0.359 & 0.358
& 0.088 & 0.087 & 0.088 \\
Week 12--26
& 0.080 & 0.094 & 0.092
& 0.045 & 0.037 & 0.040 \\
Week 26--52
& 0.057 & 0.083 & 0.083
& 0.012 & 0.009 & 0.009 \\
Week 0--52
& 0.135 & 0.152 & 0.149
& 0.039 & 0.036 & 0.036 \\
\bottomrule
\end{tabular}
\end{subtable}
\vspace{1.5mm}
\begin{minipage}{0.98\linewidth}
\footnotesize
\textit{Obs} = observed event rate (events per patient-year); 
\textit{G-comp} = G-computation estimator; 
\textit{AIPW} = augmented inverse propensity weighted estimator.
\end{minipage}
\end{table}



\end{document}